\documentclass[fleqn,10pt]{wlscirep}
\usepackage[utf8]{inputenc}
\usepackage[T1]{fontenc}
\usepackage{multicol}
\title{Short-range magnetic order and planar anisotropy in the topological ferrimagnet Mn$_3$Si$_2$Te$_6$}

\author[1,*]{Raju Baral}
\author[2]{Andrew F. May}
\author[1,+]{Stuart Calder}
\affil[1]{Neutron Scattering Division, Oak Ridge National Laboratory, Oak Ridge, Tennessee 37831, USA}
\affil[2]{Materials Science and Technology Division, Oak Ridge National Laboratory, Oak Ridge, Tennessee 37831, USA}

\affil[*]{baralr@ornl.gov}

\affil[+]{caldersa@ornl.gov}

\begin{abstract}
Mn$_{3}$Si$_{2}$Te$_{6}$ is a ferrimagnetic topological nodal-line semiconductor that exhibits unconventional colossal magnetoresitance (CMR) behavior, with short-range spin fluctuations being potentially intimately linked to the emergent properties. In this work, we determine the short range magnetic order and quantify the local magnetic anisotropy through total neutron scattering and polarized neutron powder diffraction (pNPD) measurements on polycrystalline Mn$_3$Si$_2$Te$_6$. The real space local and long range spin structure was determined through the application of magnetic pair distribution function (mPDF) analysis, with measurements from the low temperature ordered phase to the high temperature paramagnetic state. Short-range order over a frustrated trimer of three nearest neighbors was found to exist well above the long range ferrimagnetic transition. pNPD measurements in the spin polarized paramagnetic state were used to extract the local site susceptibility tensor of the Mn ions to quantify the magnetic anisotropy. Our combined mPDF and pNPD results provide quantitative information on the short-range order intrinsic to Mn$_3$Si$_2$Te$_6$, showing strong in-plane anisotropy with the spins largely confined to the $ab$-plane in zero field and remain stable with increasing temperature through the long-range to short-range ordered transition.
\end{abstract}

\begin{document}
\twocolumn

\textit{This manuscript has been authored by UT-Battelle, LLC under Contract No. DE-AC05-00OR22725 with the U.S. Department of Energy. The United States Government retains and the publisher, by accepting the article for publication, acknowledges that the United States Government retains a non-exclusive, paidup, irrevocable, world-wide license to publish or reproduce the published form of this manuscript, or allow others to do so, for United States Government purposes. The Department of Energy will provide public access to these results of federally sponsored research in accordance with the DOE Public Access Plan \\
(http://energy.gov/downloads/doepublic-access-plan).}\clearpage

\flushbottom
\maketitle
%
%
\thispagestyle{empty}

\def\TN{$T_\mathrm{N}$}
\def\Rw{$R_w$}
\def\muSR{$\mu$SR}
\def\muB{$\mu_{\mathrm{B}}$}
\def\deltampdf{3D-$\Delta$mPDF}
\def\xic{$\xi_c$}
\def\xiab{$\xi_{ab}$}
\def\MST{Mn$_{3}$Si$_{2}Te$_{6}$}
\def\An{$\mathrm{\AA}$}
\newcommand{\ra}[1]{\renewcommand{\arraystretch}{#1}}

\section*{Introduction}	

\begin{figure*}
    \centering
\includegraphics[width=0.9\linewidth]{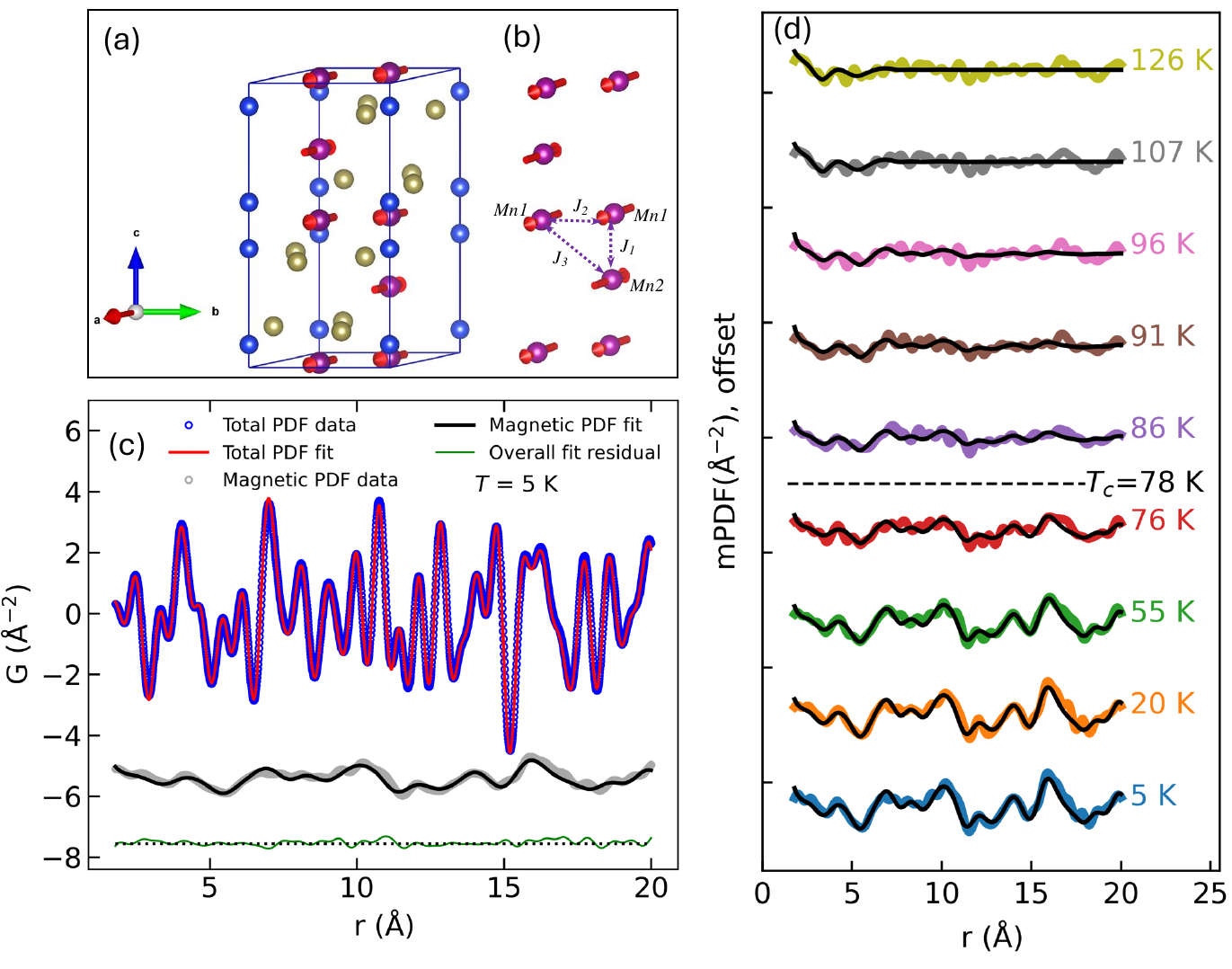}
	\caption{\label{fig:mPDF_5K} (a) Crystal structure of Mn$_{3}$Si$_{2}$Te$_{6}$ with Mn spins pointing along the {\it a}-axis. Purple, blue, and golden spheres represent Mn, Te, and Si atoms. (b) $J_1$, $J_2$ and $J_3$ represent the first three magnetic exchange interactions. (c) The neutron total PDF and isolated mPDF of Mn$_{3}$Si$_{2}$Te$_{6}$ at 5 K. The blue circles represent the total PDF data and the red curve is the total PDF fit, which is the sum of the calculated atomic and calculated magnetic fits. The gray circle and black curve represent the isolated mPDF data and fit (vertically offset for clarity), and the green symbol represents the overall fit residual. (d) Temperature series of mPDF data and fits. The colored curves represent mPDF data and the black curve represents the mPDF fit (vertically offset for clarity). The long-range order transition temperature ($ T_c=78$ K) is shown by the black horizontal dashed line.
	}		
\end{figure*}

Topological magnetic materials provide a new platform to develop magnetoelectric, transport and spintronic based devices, where the interplay between symmetry-protected electronic bands and magnetism  leads to non-trivial behavior, with the quantum anomalous Hall effect proving a well-studied example \cite{WOS:000540316700009, WOS:000763605400009, doi:10.1021/jacs.3c04249, doi:10.1021/acsmaterialsau.4c00114}. Topological nodal-line semiconductors provide a class of materials of particular interest since they offer routes to magnify unconventional behavior due to strong spin dependent Berry curvature, which can lead to enhanced magnetotransport phenomena \cite{WOS:000722366200015, WOS:000630054600001, WOS:000797036000007}. In this context, Mn$_{3}$Si$_{2}$Te$_{6}$ has been shown to be a nodal-line ferrimagnetic semiconductivity with unconventional colossal magnetoresistance (CMR), prompting intense interest \cite{Djurdji2023SRO, Ni2021,li2023intrinsically, tan2024, susilo2024high, lovesey2023anapole, zhang2022control,zhang2024variation,Zhang2023, Yin2024,wang2022,Liu2021,olmos2023pressure, PhysRevB.108.125103, NatComm_s41467-024-48432-9, NatComm_s41467-024-52350-1}. The CMR mechanism does not fall into the established category found in perovskite manganites with Jahn-Teller and double-exchange interactions \cite{WOS:A1997XZ43100005}, but instead may be closer related to the magnetic polaron behavior found in the frustrated pyrochlore manganites Tl$_2$Mn$_2$O$_7$ \cite{PhysRevLett.81.1314}. Indeed a new quantum state has been proposed in Mn$_{3}$Si$_{2}$Te$_{6}$ with predictions of chiral orbital currents being related to the unconventional CMR  \cite{zhang2022control}.

Magnetic and electronic degrees of freedom are expected to be intertwined in nodal-line topological materials. For Mn$_{3}$Si$_{2}$Te$_{6}$ there are indications of coupling between resistivity and short-range spin ordering, however the nature of the short-range magnetic ordering and how it evolves with temperature remains unknown \cite{andrew2017, Feng2022, NatComm_s41467-024-48432-9, NatComm_s41467-024-52350-1}. This forms the motivation for the present study. In general, magnetic and structural short-range order can play a significant role in determining various functional properties of materials, as evidenced by growing examples that include the paramagnon drag effect and giant spontaneous magnetostriction in MnTe \cite{baral2022realspace} \cite{baral2023giant}, negative thermal expansion in ZrW$_2$O$_8$ \cite{tucker2017ZrW2O8} and the efficiency of Li-ion batteries \cite{wang2017}. This motivates investigations to quantify and control local atomic and magnetic order in materials expected to exhibit functionally useful behavior.

Structurally, Mn$_{3}$Si$_{2}$Te$_{6}$ has alternating layers of crystallographically distinct Mn sites (Mn1 and Mn2) within the trigonal space group $P\overline{3}1c$ that form a trimer-honeycomb lattice. In this way the material can act as a bridge between 3D and 2D van der Waals materials. The Mn1 sites have twice the multiplicity of the Mn2 sites and this creates ferrimagnetic ordering, with ferromagnetic (FM) layers being coupled by antiferromagnetically alinged Mn ions within the van der Waals gap (see Fig.~\ref{fig:mPDF_5K}(a)). The magnetic ordering occurs at $T_c$=78 K with the magnetic moment predominantly aligned in the basal plane of the trigonal structure, however, any out of plane $c$-axis component is potentially linked to the field induced CMR \cite{andrew2017, zhang2022control, Feng2022}. 

Considering the magnetic interactions, Mn$_{3}$Si$_{2}$Te$_{6}$ is locally frustrated, with a trimer of nearest neighbor magnetic exchange interactions all being antiferromagnetic (AFM) despite the moments in the trimer forming both AFM and FM alignment. The spin alignment and exchange interactions are shown in Fig.~\ref{fig:mPDF_5K}(b), with the three nearest neighbor trimer highlighted. Recent experiments and theoretical calculations of the exchange interactions have shown that the nearest neighbor $J_1$ (Mn1-Mn2) is the largest interaction, however the third nearest neighbor interaction $J_3$ (Mn2-Mn1), dominates over the in-plane $J_2$ (Mn1-Mn1) interaction \cite{andrew2017, salaMnSiTe}. This AFM $J_2$ exchange interaction is frustrated and instead forms FM aligned moments to produce the emergent ferrimagnetic long range order ground state. The exchange interactions lead to a competition between an antiferromagnetic and ferrimagnetic ground state, which are close in energy and results in a relative suppression of the magnetic ordering temperature \cite{andrew2017, Zhang2023}. This suppression of long range ordering temperature from magnetic frustration, along with the layered structure, provide mechanisms for the observed short range order from previous neutron scattering measurements at 100 K \cite{Feng2022} and 120 K \cite{andrew2017}. A quantitative analysis of such diffuse scattering, as well as following the evolution into the long range ordered state, is non-trivial and has not been performed for Mn$_{3}$Si$_{2}$Te$_{6}$.

In this work, a comprehensive view and quantification of the short-range magnetic order and local magnetic anisotropy in Mn$_{3}$Si$_{2}$Te$_{6}$ is provided through neutron total scattering and polarized neutron powder diffraction (pNPD) measurements. By applying magnetic pair distribution function (mPDF) analysis \cite{frandsen2015mPDFMnO,frandsen2022diffpympdf} of the total neutron scattering data, a real-space visualization of the local spin-spin correlations is presented, both in the ordered magnetic state and in the paramagnetic regime. The measurements reveal magnetic diffuse scattering that persists well above the transition temperature, and these are successfully model directly in real space. Polarized neutron scattering measurements in the paramagnetic state provide a measurement of the local site susceptibility tensor to quantify and visualize the local magnetic anisotropy at the Mn sites. Collectively the results indicate strong in-plane anisotropy and the implications of this above $ T_c$ are considered on the observed physical properties of Mn$_{3}$Si$_{2}$Te$_{6}$.

\section*{Results}

\subsection*{Short-range order in Mn$_{3}$Si$_{2}$Te$_{6}$}

To explore the local magnetic order in Mn$_{3}$Si$_{2}$Te$_{6}$ an analysis of the atomic PDF and mPDF data for temperatures spanning well above and below $ T_c$ =78 K was performed. We begin with the low temperature magnetic ordered phase at 5 K. Fig.~\ref{fig:mPDF_5K}(c) represents the atomic and magnetic fits of the total neutron scattering data in real space using a local fitting range of 1.8 - 20 \AA. The total neutron scattering data includes contributions from both atomic and magnetic scattering, so the magnetic signal has to be isolated. This was done in real space.  To isolate the mPDF signal, we first performed the atomic PDF fit using the known atomic structure (space group $P\bar{3}1c$) and calculated the atomic PDF scattering. The fit residual from this fit contains the magnetic signal which is then treated as the magnetic PDF data to perform the mPDF refinement. Details of the atomic and magnetic fitting approach in real space are discussed in Refs.\cite{frandsen2015mPDFMnO,baral2022realspace,baral2024localSpinStructure}. The low goodness of fit, $R_{w}$=0.0535, and overall fit residual shown by the green curve indicates a close match between the measurement and model. The orientation of the local magnetic moment is characterized in a spherical coordinate system by a polar angle ($\theta$) and an azimuthal angle ($\phi$). $\theta$ specifies the spin direction relative to the crystallographic $c$-axis, while $\phi$ defines the spin direction relative to the crystallographic $a$-axis. The in-plane angle does not change the long range or local mPDF pattern and was fixed to $\phi=0^\circ$, whereas the polar angle was free to refine. The best magnetic fit occurs when the Mn spins are only in the $ab$-plane ($\theta = 90^\circ$ with the $c$-axis) as shown in Fig.\ref{fig:mPDF_5K} (a), which is equivalent to previous studies \cite{andrew2017}. Further consideration of any $c$-axis component is discussed later.

Atomic and magnetic PDF refinements were also performed using the lower-symmetry space group $C2/c$. mPDF refinements yielded polar angles of Mn1 and Mn2 spins of $85^\circ$ from the $c$-axis and the azimuthal angle of the Mn1 spin was refined to $178^\circ$ from the $a$-axis, while the azimuthal angle of Mn2 was fixed to zero due to magnetic symmetry constraints. However, this did not produce any significant change in the mPDF pattern compared to spins fixed to the $ab$-plane. Consequently, the higher-symmetry space group is used for the subsequent analysis. A comparison of fits for the two space groups is shown in the supplementary section. To obtain reliable refinements the moments of Mn1 and Mn2 were constrained to be collinear and fixed along the $a$-axis, as was done in Ref.~\cite{andrew2017} and for applied field measurements in Ref.~\cite{Feng2022}.

\begin{figure*}
    \centering
	\includegraphics[width=0.65\linewidth]{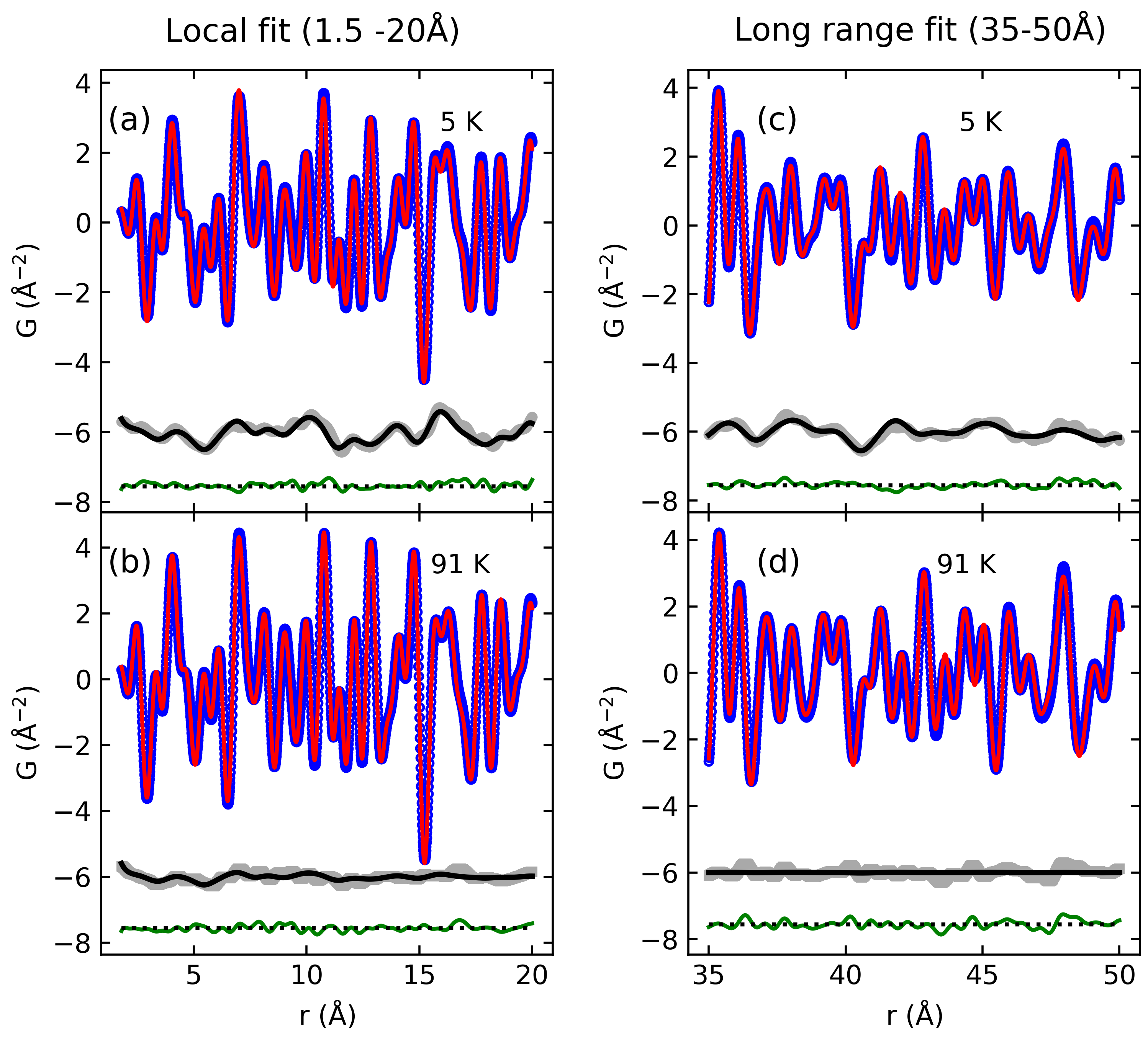}
	\caption{\label{fig:lowr_highr_fit}  Atomic and magnetic PDF fits of Mn$_{3}$Si$_{2}$Te$_{6}$ for two different fitting ranges; a local fit (1.8 - 20 \AA) and a long range fit (35 - 50 \AA) at 5 K and 91 K.
	}		
\end{figure*}

\begin{figure*}
    \centering
	\includegraphics[width=0.9\linewidth]{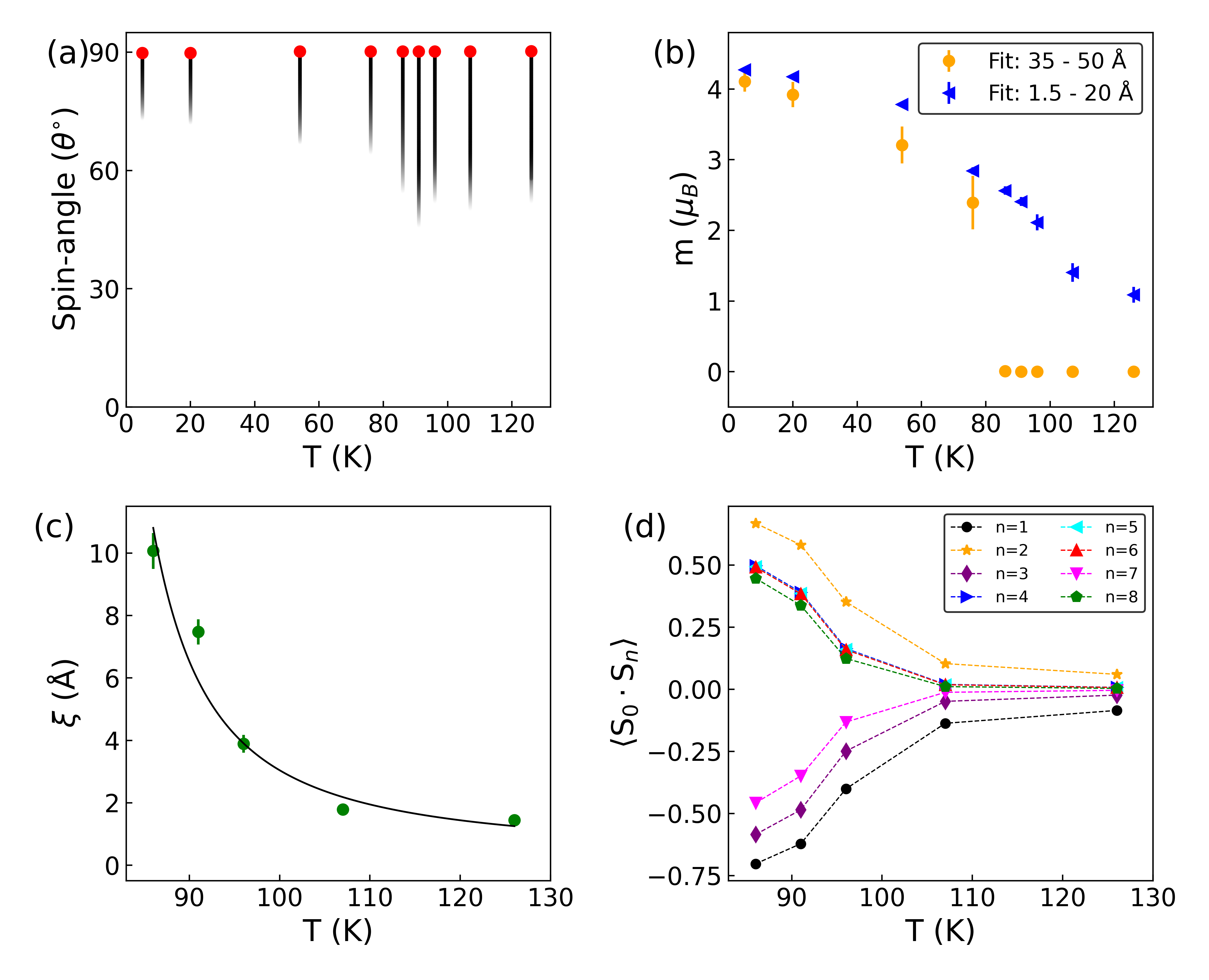}
	\caption{\label{fig:m_xi}  (a) Best fit spin angle (red circle) as a function of temperature calculated over a fitting range 1.8 - 20 \AA. The spin angle indicates the angle between the direction of the ordered magnetic moment and $c$-axis of the unit cell. The shaded bars represent a confidence interval described in the text.    
    (b) The ordered magnetic moment ($m$) of Mn$_{3}$Si$_{2}$Te$_{6}$ as a function of temperature extracted from mPDF fits from both a local fitting range of 1.8 - 20 \AA~(blue symbols)  and long range fitting range of 35 - 50 \AA~(orange symbol). (c) Isotropic correlation length (green symbols) as a function of temperature in above $T_c=78$ K, calculated over a local fitting range of 1.8 - 20 \AA. (d) Average spin correlation function $\langle \mathrm{S}_0 \cdot \mathrm{S}_n \rangle$ with temperature for the first nine nearest neighbors in the paramagnetic regime.
	}		
\end{figure*}

Having established the mPDF model at 5 K, we now extend this with a series of atomic and magnetic PDF refinements for data collected across the temperature range 5 to 126 K. This was performed with the same local real space fitting range of 1.8 - 20 \AA, as shown in Fig. \ref{fig:mPDF_5K}(d). Approaching $ T_c=78$ K in the long-range ordered regime the magnetic model captures the mPDF data well. Above $ T_c$ in the paramagnetic regime at 86 K prominent diffuse scattering is observed, indicating the persistence of short-range magnetic order extending up to 20 \AA. The diffuse scattering above 86 K weakens gradually with increasing temperature as a result of thermal fluctuations and the scattering is reduced to a smaller and smaller real-space region. However, clear features of short-range magnetic ordering are present even at 126 K, which is the maximum temperature measured for this experiment. The reciprocal space neutron data before transformation to real space can be viewed in the supplemental information and, as expected, shows such diffuse scattering.

A series of fits were performed to determine the best-fit spin direction at each temperature from the mPDF calculations. For each temperature, the angle between the spin direction and $c$-axis was varied from 0$^\circ$ to 90$^\circ$ in increments of 5$^\circ$. The goodness of fit metric ($R_w$) was computed for each temperature and the angle corresponding to the lowest $R_w$ was identified as the best-fit angle. These best-fit angles are shown by the red circle, while the black shaded bars underneath indicate the confidence interval for the best-fit spin angles (see Fig.~\ref{fig:m_xi}(a)). The shading intensity of the bars lightens progressively as $R_w$ increases from the optimal value, fading to white at a threshold of 0.5\% and above. A threshold of 0.5 \% was chosen as the point that makes a meaningful difference between the fits. Fig. \ref{fig:m_xi}(a) represents the variation of the spin-angle with temperature. The spin angle remains constant with increasing temperature, with spins confined to the $ab$-plane ($\theta=$90$^{\circ}$ to the $c$-axis). 

To consider the spin ordering further and distinguish between local and long range order, mPDF fits were performed over two different fitting ranges; one with a fitting range of 1.8 - 20 \AA (local magnetic order) and the other with a fitting range of 35 - 50 \AA~ (average long range magnetic order). Fig.~\ref{fig:lowr_highr_fit}(a, b) shows the combined atomic and mPDF refinements for the two different fitting ranges at temperatures of 5 K and 91 K. At 5 K, both fitting ranges captured the mPDF patterns as expected, as the material is inside a long-range order regime. At 91 K, however, there is a clear difference in the mPDF pattern calculated using these two different fitting ranges. The local mPDF fit (1.8 – 20 \AA) successfully captured the observed short-range magnetic correlations, whereas the average long-range fit (35–50 \AA) failed to do so. The mPDF data for the average long range fit shows no discernible features, resulting in a flat mPDF fit, indicating that the average long-range fit is not sensitive to the short-range magnetic correlations of the material. 

The mPDF refinements for the temperature series allow the extraction of the magnetic moment, correlation length, and spin-angle of the moments over a defined real-space fitting range. The magnetic moments extracted using the different local (1.8 - 20 \AA) and long range (35 - 50 \AA) fitting regions are shown in Fig.~\ref{fig:m_xi}(b). Moments extracted using the average long-range fit are essentially zero at $T>T_c$. However, a sharp increase in moment occurs for $T<T_c$ indicating a quick onset of long-range magnetic order. The moments extracted using the local fit decrease almost uniformly with increasing temperature through $T_c$ and remain non-zero even at 126 K, indicating the persistence of short range order well above the magnetic transition.

Similarly, Fig.~\ref{fig:m_xi}(c) shows the temperature dependency of the isotropic correlation length for $T>T_c$. These correlation lengths were calculated by implementing the exponential envelope ($e^{-r/\xi}$) in the mPDF calculation, using $\xi$ as the fitting parameter. For temperatures $T>T_c$, the isotropic correlation length decreases with increasing temperature. The correlation length below $T_c$, however, is infinite since magnetic order is long range.

To investigate the behavior of the correlation length in the paramagnetic regime, we performed a fit based on the  Berezinskii-Kosterlitz-Thouless (BKT) prediction \cite{Kosterlitz_1973}, described by the equation:
\begin{eqnarray}
    \xi (T) = A \cdot e^{b \left(\frac{T_{KT}}{T - T_{KT}}\right)^{1/2}},
\end{eqnarray}

where $A$ and $b$ are non-universal constants. This was proposed to describe the phase transition of 2D XY model materials where spins are largely confined to be in the plane. The XY model has been shown to be applicable to Mn$_3$Si$_2$Te$_6$ from theoretical calculations and measurements \cite{salaMnSiTe, Zhang2023}. Considering the temperature evolution of the correlation length, Fig.~\ref{fig:m_xi}(c), the best fit was obtained with the parameters, $A=0.119 \pm 0.056$, $b=2.067 \pm 0.485$, and $ T_{KT}=71 \pm 3$ K. The ratio $T_{KT}/T_c=71/78=0.91$, aligns with typical values for materials exhibiting 2D properties that are candidates for BKT behavior.

\begin{figure*}
    \centering
	\includegraphics[width=0.85\linewidth]{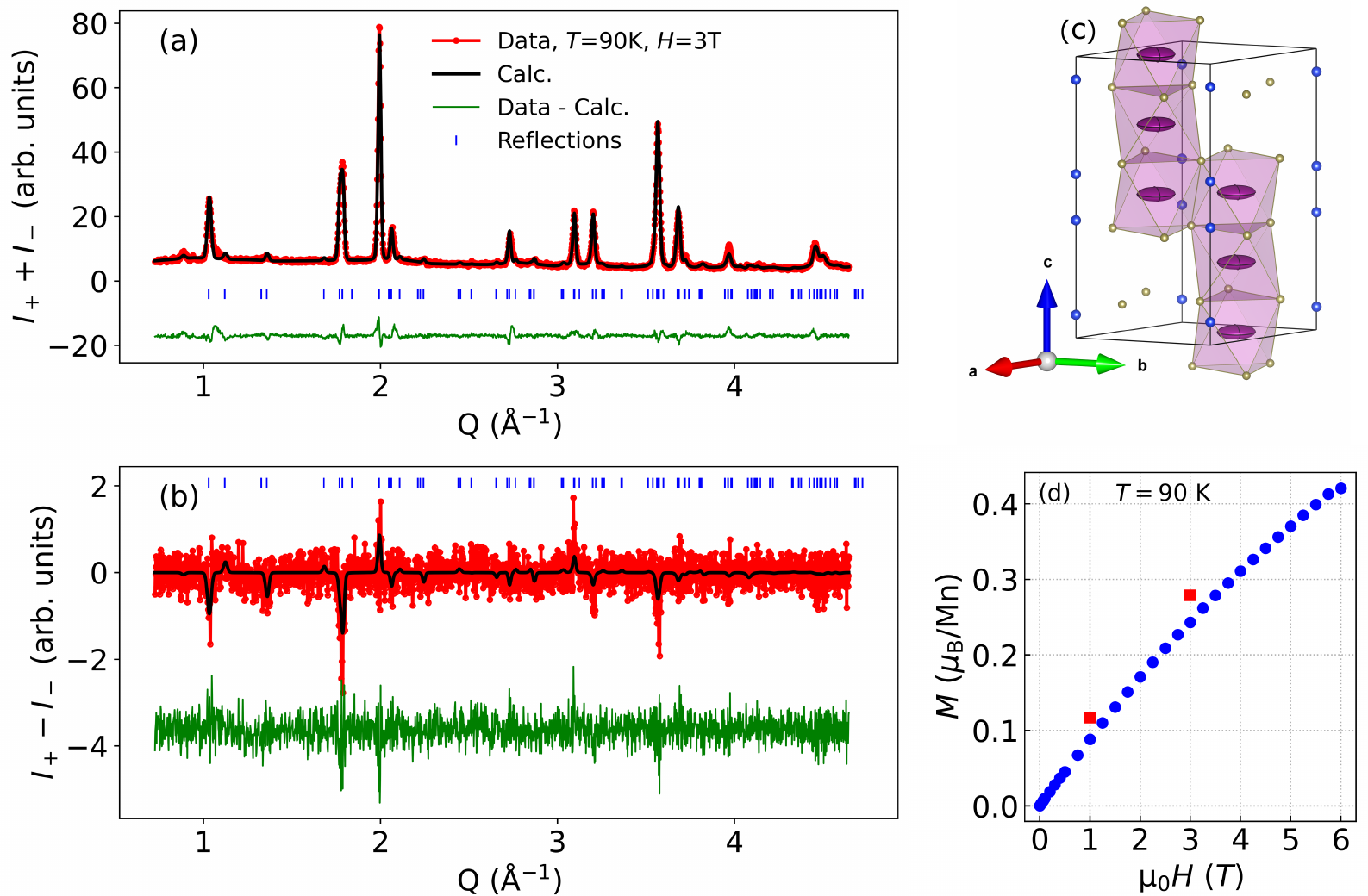}
	\caption{\label{fig:polarized_90K_3T} Data and fits of the polarized neutron powder diffraction (pNPD) measurements for the (a) sum ($I_{+} + I_{-}$) and (b) difference ($I_{+} - I_{-}$) at 90 K (paramagnetic regime) under an applied magnetic field of 3 T. Red, black, and green symbols represent the data, fit, and fit residual, respectively. The vertical blue lines represent the Bragg reflection positions. (c) Magnetization ellipsoids (purple) for Mn are shown within the unit cell. (d) Isothermal magnetization on a polycrystalline sample of Mn$_3$Si$_2$Te$_6$ at 90 K (blue circles). The red squares show the value of the magnetization at 1 T and 3 T obtained from the pNPD calculations. 
    }
\end{figure*}

The average spin correlation function ($\langle \mathrm{S}_0 \cdot \mathrm{S}_n \rangle$) for the first eight coordination shells were calculated using mPDF as a function of temperature in the paramagnetic regime. Fig.~\ref{fig:m_xi}(d) shows the variation of the average spin correlation function $\langle \mathrm{S}_0 \cdot \mathrm{S}_n \rangle$ with temperature, where S$_0$ is an arbitrary spin, S$_n$ is a spin in the $n^{th}$ coordination shell relative to S$_0$, and angle brackets represent an average overall S$_n$ in the shell. The spins in the calculation have unit length. Negative and positive values correspond to net AFM and FM spin directions, respectively. At 86 K, the first nearest neighbor ($n=1$) correlation is AFM, and the second nearest neighbor ($n=2$) correlation is FM, and the third nearest neighbor ($n=3$) is again AFM and so on. As expected the spin correlation function decreases with a rise in temperature, however, even at 107 K and 126 K  first nearest neighbor, second nearest neighbor and third nearest neighbor correlations are evident and can be identified as AFM, FM and AFM. This highlights the persistence of the ferrimagnetic ordering of Mn-Mn ions, even on the local level above $ T_c$. We emphasize that these are the spin-spin correlations that result from the competing magnetic exchange interactions, which are all antiferromagnetic in nature.

\subsection*{Local susceptibility tensor in Mn$_{3}$Si$_{2}$Te$_{6}$}

To gain a further understanding of the local anisotropy of Mn, which is central to the emergent magnetic behavior, we conducted polarized neutron powder diffraction (pNPD) measurements. This technique is described in detail in Refs.~\cite{Gukasov_2010, kibalin2019local}. The local site susceptibility tensor was extracted from these measurements by applying a magnetic field to the sample and measuring the diffraction pattern obtained with the polarization of the incident neutron beam parallel ($I_+$) or antiparallel ($I_{-}$) to the field direction. Fig. \ref{fig:polarized_90K_3T}(a, b) shows the measured and calculated sum ($I_+ + I_{-}$) and difference ($I_+ - I_{-}$) diffraction patterns collected on Mn$_3$Si$_2$Te$_6$ at 90 K in a magnetic field of 3 T.

The information accessed in pNPD measurements can be found by considering the intensity in a neutron measurement, which  has  the form:
\begin{align}
I &\propto F_N^2 + {\vec F}_{M\perp}^2 +  {\vec P}\cdot(F_N^*{\vec F_{M\perp}} + F_N{\vec  F_{M\perp}^*}) 
\end{align}

where $F_N$ and ${\vec F_M}$ are the nuclear and magnetic structure factors and $\vec{P}$ is the polarization value of the incident beam. An additional chiral term is neglected in equation (2) since this is not accessed in powder measurements. In unpolarized neutron measurement only the first two terms are measured, with the third term zero. This third term is the magnetic-nuclear interference term and if the polarization state of the neutron beam is known and controlled, it can be accessed. The flipping difference method is used, with the intensity difference of spin up ($I_+$)  and spin down ($I_-$) pNPD measurements given by:
\begin{eqnarray}
\Delta I = I_+ - I_- \propto  2(F_N^\star \langle {\vec F}_{M,\perp}  \cdot  {\vec P} \rangle  + F_N \langle {\vec F}_{M,\perp}^\star  \cdot  {\vec P} \rangle  )  
\label{equation:Polarized_diff} 
\end{eqnarray}

Angular brackets account for the powder averaging. The magnetic structure factor is given by ${\vec F}_{M}(Q)=\sum_i {\vec m_i}f_m(Q) \exp(iQ.r_i)$, where the sum is over the unit cell, $f_m(Q)$ is the magnetic form factor and ${\vec m_i}$ is the magnetic moment on atom $i$. In unpolarized measurements ${\vec m_i}$ is considered as the local moment on the magnet site on atom $i$ that forms long range or short range magnetic order. In the local site susceptibility tensor approach ${\vec m_i}$ is the induced magnetic moment on atom $i$ by a magnetic field. ${\vec m_i}$ in a magnetic field ${{\vec H}_j}$ satisfies the behavior:
\begin{eqnarray}
\hat{\chi}_{ij}  = 
\begin{pmatrix}
{\chi_{11}} & {\chi_{12}} & {\chi_{13}}\\
{\chi_{12}} & {\chi_{22}} & {\chi_{23}}\\
{\chi_{13}} & {\chi_{23}} & {\chi_{33}}\\
\end{pmatrix}
=\frac{{\vec m_i}}{{\vec H}_j},
\label{sus_tensor}
\end{eqnarray}

The magnetic scattering can now be described as a structure factor tensor:
\begin{eqnarray}
 {\vec F}_{M,\perp}(Q)=\sum_i {\vec \chi_{ij}}f_m(Q) \exp(iQ.r_i)\cdot {\vec H}
\label{equation:Polarized_magn} 
\end{eqnarray}

The site symmetry determines any constraints between matrix elements and the allowed non-zero tensor components $\chi_{ij}$. The refinements were performed for the trigonal space group $P\bar{3}1c$ that is applicable in the paramagnetic region in which the measurements were undertaken. For this symmetry only the diagonal terms in the tensor are allowed. We found no improvement in the fitted results if the lower monoclinic symmetry $C2/c$ was used, however this would confine the magnetization to be uniquely along the $a$-axis for Mn1 rather than unconstrained within the $ab$-plane. The CrysPy software was used to extract $\bar{\chi_i}$ directly from pNPD measurements \cite{cryspy}. The local site susceptibility tensors were refined against both the sum and difference data, as shown in Fig.~\ref{fig:polarized_90K_3T}(a,b). Allowing the two different Mn sites to refine separately produced the same result as constraining them together. The susceptibility tensors for the 90 K measurements in an external field of 3 T for the Mn ions was found to be:

\begin{center} 
$\begin{pmatrix}
		0.28(3) & 0        & 0\\
			0   & 0.28(3)  & 0\\
			0   & 0        & 0.09(5)
	\end{pmatrix}$ $\mu_B$/T\\
\end{center}

The Mn anisotropy can be visualized in Fig.~\ref{fig:polarized_90K_3T}(c) as magnetization ellipsoids. These results show the strong local planar anisotropy within Mn$_{3}$Si$_{2}$Te$_{6}$ that drives the magnetic ordering of spins primarily constrained in the $ab$-plane, consistent with the mPDF results.

The powder average magnetization determined from the pNPD measurements is 0.28 $\mu_B$/Mn, which is in very close agreement with the value of 0.24 $\mu_B$/Mn obtained from magnetization measurements on a polycrystalline sample at 90 K and 3 T, similar agreement was found for measurements at 1 T and 90 K, as shown in Fig.~\ref{fig:polarized_90K_3T}(d). This consistency confirms the robustness and reliability of the pNPD analysis.

\section*{Discussion}

The short-range magnetic behavior of the  topological ferrimagnetic material Mn\(_{3}\)Si\(_{2}\)Te\(_{6}\) investigated using total neutron scattering mPDF analysis revealed short-range magnetic order exists well above $T_c$. Previous measurements found signatures of local magnetic behavior \cite{andrew2017, Feng2022}, but the nature of the ordering has remained an open question. This incomplete understanding of the magnetic ordering hamstrings further investigations on  Mn\(_{3}\)Si\(_{2}\)Te\(_{6}\), particularly with suggestions that the CMR and the local ordering could be coupled \cite{Feng2022}. Here, the total neutron data were modeled effectively in real space and showed ordering over multiple spin-spin distances that persisted to a trimer of three nearest neighbors at the highest measured temperatures. This trimer contains two Mn1 sites and one Mn2 site, which are key to the ferrimagnetic ordering in the material. The magnetic exchange interaction is antiferromagnetic between all Mn-Mn bonds in the trimer, despite the ordering being a mixed ferromagnetic and antiferromagnetic spin state in the trimer. The robustness of this frustrated trimer may be a key ingredient in the CMR and related reports of potential chiral orbital currents \cite{zhang2022control}.

Considering the magnetic anisotropy, the pNPD results at 90 K highlighted strong planar XY anisotropy with a determination of the local site susceptibility tensor for the Mn ions. This points to the importance of spin-orbit coupling and non-negligible orbital moment to produce this anisotropy. The unconventional CMR that occurs in  Mn\(_{3}\)Si\(_{2}\)Te\(_{6}\) is driven by an applied field canting the spins out of the $ab$-plane. The results here show that there is a sizeable anisotropy that is required to be overcome to allow such a canting. Moreover, it indicates a similar anisotropy on both Mn1 and Mn2 magnetic ions, despite their distinct crystallographic environments.

The details of the spin directions in  Mn\(_{3}\)Si\(_{2}\)Te\(_{6}\) over a wide temperature range were extracted using mPDF. The spins were found to remain principally fixed in the $ab$-plane at all temperatures, both within the long range ordered transition and at higher temperatures in the short-range ordered temperature range. Indeed a smooth variation, with no apparent abrupt change at $T_C$, was observed in the ordering on a local level. Given the apparent link between the CMR and magnetic ordering, this indicates that the conditions for the CMR in Mn\(_{3}\)Si\(_{2}\)Te\(_{6}\) may be in place well above the long range magnetic order transition.

The pNPD and mPDF results presented support a 2D XY Hesienberg model for  Mn\(_{3}\)Si\(_{2}\)Te\(_{6}\). A consideration of the behavior within a BKT transition, where planar anisotropic behavior is applicable, was undertaken. Formally the BKT transition describes a topological transition from unpaired vortices to bound vortex-antivortex pairs at a critical temperature $T_{KT}$ and has been applied to transitions in materials such as MnPS$_3$ \cite{RONNOW2000676}, BaNi$_2M_2$O$_8$ (M=V,P) \cite{PhysRevLett.91.137601, 10.1063/1.348816} and $A$Cr$_2$O$_4$ ($A$=Mg,Zn,Cd) \cite{PhysRevB.95.224101}. Indeed it was proposed that the BKT transition could be a mechanism to drive colossal magnetoresistance \cite{PhysRevB.104.L020408}, making it of potential interest for the behavior in Mn$_3$Si$_2$Te$_6$. The spin correlation extracted from the mPDF analysis in Mn\(_{3}\)Si\(_{2}\)Te\(_{6}\) above $ T_c$ follow this model to a good approximation. This would support a XY Heisenberg interpretation of Mn$_{3}$Si$_{2}$Te$_{6}$ and may have implications on the mechanism driving the colossal magnetoresistance and the link to short-range magnetic order. We emphasize that this is just one of several potential models that could be applied to the short-range behavior. Future experimental and theoretical investigations on the role of the KBT transition in Mn$_3$Si$_2$Te$_6$ will be of interest, as well as implementations of further models motivated by the details of the local ordering uncovered in the results presented.

In summary, in Mn\(_{3}\)Si\(_{2}\)Te\(_{6}\) the length scales, magnetic moment sizes, anisotropy and spin directions of the short-range magnetic ordering have been determined. The local exchange frustration in the Mn trimer fixes the spin ordering above $T_C$ to produce ferrimagnetic ordering which is largely confined to the $ab$-plane due to spin-orbit coupling, but with an allowed component along the $c$-axis. This ordering remains robust as the material undergoes long range magnetic order, in an apparent continuous transition as the length scales of the ordering increases from the single magnetic trimer to infinitely long-ranged. The effect of perturbing the local Mn environment, through chemical disorder or external perturbations, may offer promising future avenues for investigations in this material as a way to control the unconventional CMR and related behavior by removing the frustration in the exchange interactions and consequently altering the local magnetic order.

\section*{Methods}
\subsection*{Sample synthesis}

Polycrystalline Mn$_{3}$Si$_{2}$Te$_{6}$ was made by reacting Te, Mn, Si at a stoichiometry of Mn$_3$Si$_{2}$Te$_6$.  The elements were heated in a SiO$_2$ tube sealed under 1/5 atm of argon.  The ampoule was heated to 1100$^\circ$C for over 10 hours, then quenched into ice-water. The ampoule was then returned to the furnace at 800$^\circ$C and annealed for approximately 72 hours before the furnace was powered off.  Following this, the sample was ground in air and annealed for another 72 hours at 800$^\circ$C.  While the experiments performed do not provide conclusive evidence that excess Si is required, the impurity phases observed in initial reactions led to this approach together with the quenching from a molten phase at 1100$^\circ$C.  Generally, making pure Mn$_{3}$Si$_{2}$Te$_{6}$ powder is a somewhat difficult task due to the formation of Mn-Te binary phases.  The purity of the powder was first checked with a laboratory x-ray diffractometer. Then, the temperature-dependent magnetization was measured and confirmed to be consistent with the behavior previously observed in melt-grown Mn$_{3}$Si$_{2}$Te$_{6}$ crystals with indications of a magnetic anomaly near room temperature observed.

\subsection*{Neutron total scattering}

A neutron total scattering experiment was conducted on the HB-2A powder diffractometer at the High Flux Isotope Reactor (HFIR) in Oak Ridge National Laboratory (ORNL) \cite{garlea2010HFIR, powderHB2A2018}. HB-2A is traditionally utilized for reciprocal space studies, however recently it has been shown that total scattering mPDF studies are highly feasible \cite{baral2024localSpinStructure,BenZnMnTe}. A vertically focusing germanium monochromator was used to select the wavelength of 1.12~\AA~from the Ge(117) reflection. Approximately 7.3 grams of sample was loaded into a vanadium sample holder inside a helium glove box. The total scattering data were collected at various temperatures ranging from 5 to 126 K, with a collection time of 4 hours for each temperature point. The collected data were then normalized using a vanadium standard measurement. The detector position and the wavelength of the neutron beam were calibrated with a measurement of a Si standard. The collected $Q$-space data were further Fourier-transformed using PDFgetN3~\cite{juhas2018PDFgetN3} with a \textit{Q}$_{max}$=10 Å$^{-1}$ to produce the PDF data. The combined atomic and magnetic refinements were performed using {\it diffpy.mpdf}~\cite{frandsen2022diffpympdf}, a Python-based software designed for atomic and mPDF calculations and fitting.

\subsection*{Half-polarized neutron powder diffraction}
Half-polarized neutron powder diffraction (pNPD) measurements were performed using the same powder diffractometer, HB-2A  \cite{garlea2010HFIR, powderHB2A2018}, at HFIR in ORNL. Approximately 7.3 grams of Mn$_{3}$Si$_{2}$Te$_{6}$ powder was pressed into pellets of 8 mm diameter.  These pellets were then vertically stacked and loaded into a vanadium can, ensuring that there was no movement of the powder grains when the magnetic field was applied. This measurement was carried out in a cryomagnet using neutrons with a constant wavelength of 2.41 \AA. The incident neutron beam was polarized with a V-cavity supermirror and a guide flipper was used to control the two neutron beam spin states, spin-up and spin-down. Polarized measurements were taken with the spin-up and spin-down states of the incident neutron beam at 90 K in an applied magnetic field of 1 and 3 Tesla. Unpolarized zero field measurements at 100 K were taken as a reference for data analysis. The data analysis was performed using the open source {\it CrysPy} \cite{cryspy} software to determine the local site susceptibility tensor.

\section*{Data availability}
Data included in this study are available from the corresponding authors upon reasonable request.

\section*{Acknowledgments}
This research used resources at the High Flux Isotope Reactor and Spallation Neutron Source, a DOE Office of Science User Facility operated by the Oak Ridge National Laboratory. The beam time was allocated to HB-2A (POWDER) on proposal number IPTS-31942.1 and to NOMAD on mail-in proposal number IPTS-31797.1. Supported by the U.S. Department of Energy, Office of Science, Basic Energy Sciences, Materials Sciences and Engineering Division.

\section*{Author contributions statement}

R.B. and S.C. conducted the neutron scattering experiments and analysed the results. A.F.M. synthesized the sample and performed characterizations.  All authors reviewed the manuscript. 

\section*{Additional information}

\textbf{Competing interests:} The authors declare no competing interests.


\bibliography{bibfile.bib}

\clearpage
\onecolumn

\section*{Supplementary Note 1: Neutron diffraction data in reciprocal space}

The neutron total scattering structure function $S(Q)$ is shown for Mn$_3$Si$_2$Te$_6$ ~at various temperatures in Supplementary Figure \ref{fig:SofQ}. The structure function $S(Q)$ is displayed only up to 6~\AA$^{-1}$. The $S(Q)$ plots of Mn$_3$Si$_2$Te$_6$ show prominent magnetic Bragg peaks below $\rm T_c = 78$ K, for example around 0.88~\AA$^{-1}$, 1.36~\AA$^{-1}$, 2.24~\AA$^{-1}$, and 2.87~\AA$^{-1}$. As the temperature increases above $T_c$, the first sharp peak becomes broad and diffuse due to short-range magnetic correlations, which gradually weaken with increasing temperature.

\begin{figure*}
\setcounter{figure}{0}
\renewcommand{\figurename}{Supplementary Figure}
\renewcommand{\thefigure}{\arabic{figure}}
	\includegraphics[width=1\linewidth]{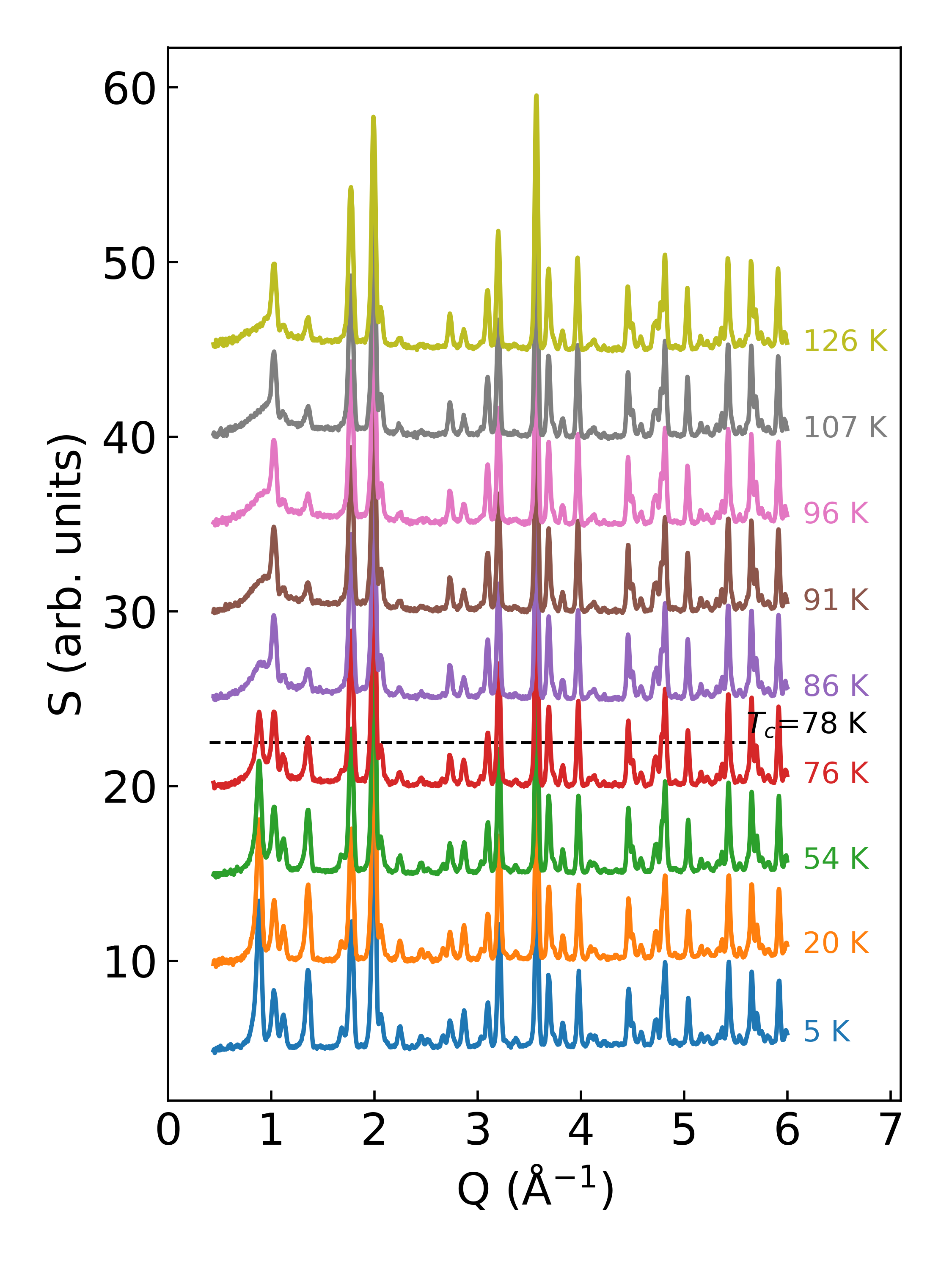}
	\caption{\label{fig:SofQ} Reciprocal space neutron diffraction data collected in the HB-2A powder diffractometer for various temperatures for Mn$_3$Si$_2$Te$_6$. Vertically offset for clarity.}
		
\end{figure*}

\section*{Supplementary Note 2: Comparison atomic and magnetic fits at 5 K using two different space-groups}

\begin{figure*}
\renewcommand{\figurename}{Supplementary Figure}
\renewcommand{\thefigure}{\arabic{figure}}
	\includegraphics[width=1\linewidth]{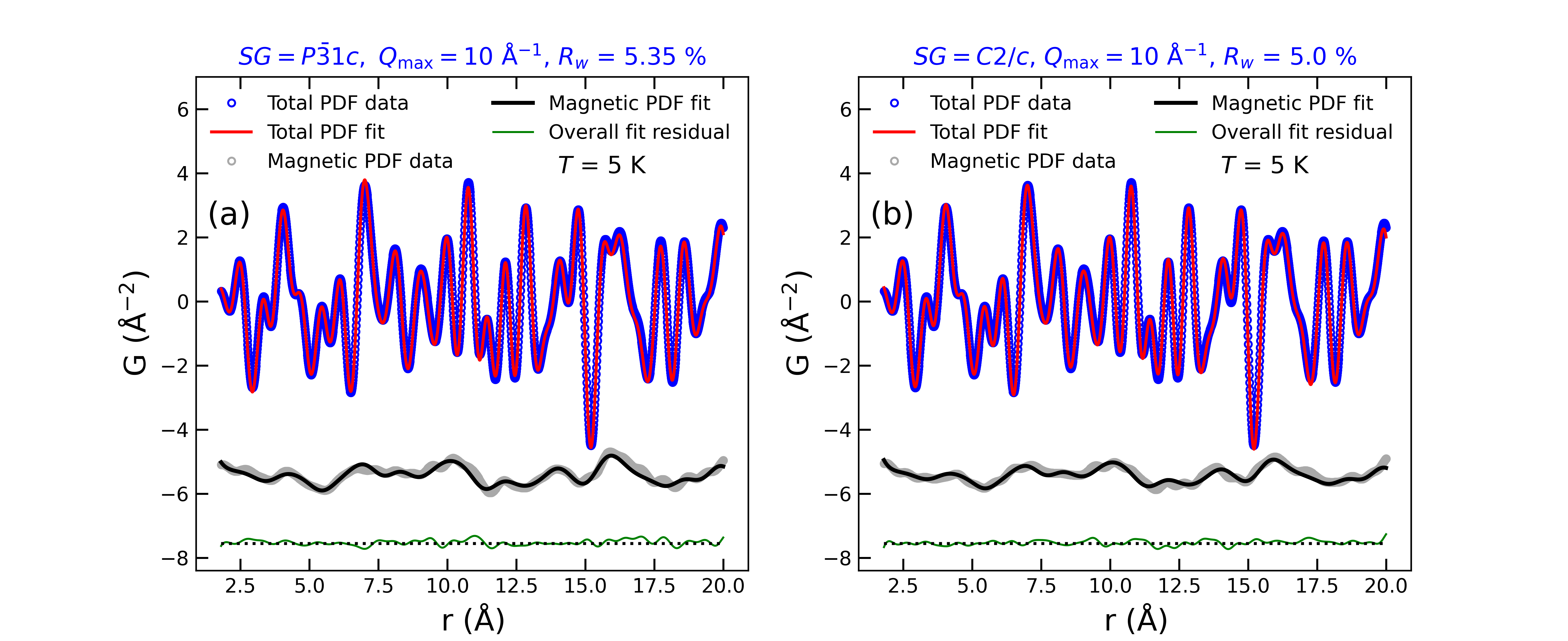}
	\caption{\label{fig:mPDF_5K_different_SG} Atomic and magnetic PDF fit at 5 K, generated using $Q_{\text{max}}= 10\ \text{\AA}^{-1}$, for two different space groups. (a) Space group $P\bar{3}1c$, and (b) Space group $C2/c$. }
\end{figure*}

Supplementary Figure \ref{fig:mPDF_5K_different_SG} shows the atomic and magnetic PDF fits for the data collected at 5K, generated with $Q_{\text{max}}= 10\ \text{\AA}^{-1}$ for two different space groups; (a) $P\bar{3}1c$ and (b) $C2/c$. Comparable fits are obtained for both space groups.



\section*{Supplementary Note 3: High temperature atomic PDF fits}

\begin{figure*}
\renewcommand{\figurename}{Supplementary Figure}
\renewcommand{\thefigure}{\arabic{figure}}
	\includegraphics[width=1\linewidth]{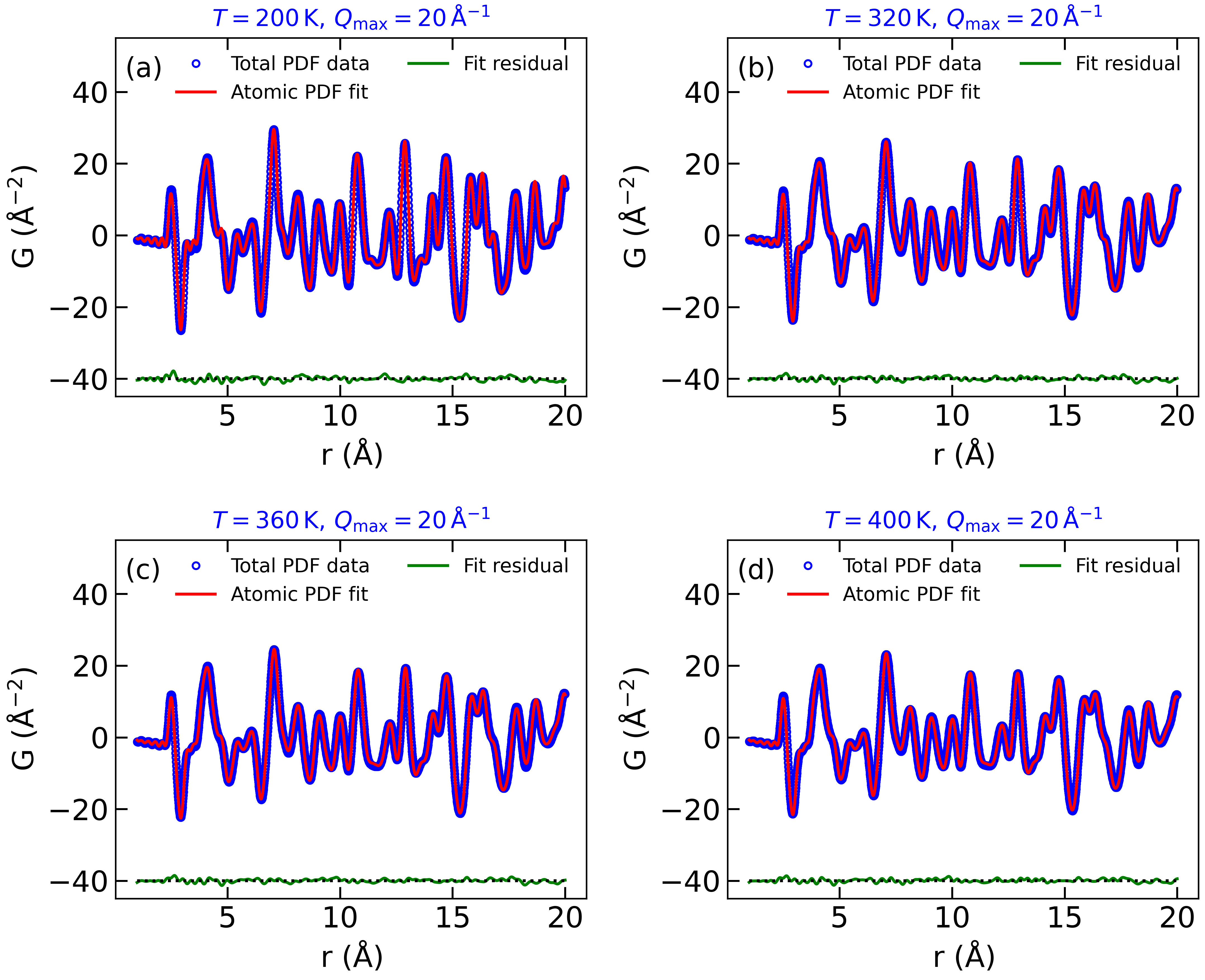}
	\caption{\label{fig:highT_PDF} Atomic PDF fits at high temperatures using $Q_{\text{max}}= 20\ \text{\AA}^{-1}$. The blue, red, and green symbols represent the total PDF data, atomic PDF fit, and fit residual, respectively.}
\end{figure*}

Supplementary Figure \ref{fig:highT_PDF}(a-d) shows the atomic PDF fits at 200, 320, 360, and 400 K, respectively, performed using the published crystallographic structure. This dataset was collected at Nano-Scale-Ordered Material diffractometer (NOMAD) at Spallation Neutron Source (SNS), Oak Ridge National Laboratory (ORNL). The total neutron scattering data were processed and transformed with $Q_{\text{max}}= 20\ \text{\AA}^{-1}$  using NOMAD's automated data reduction scripts.  The figures demonstrate a good agreement between the experimental data and the atomic model. No evidence of correlated magnetism on the local level was observed at 200 K or higher in the data collected.

\end{document}